\documentclass[11pt]{article}
\usepackage{moriond,epsfig}

\def\PRL{\em Phys. Rev. Lett.}

\def\be{\begin{equation}}
\def\ee{\end{equation}}
\def\bea{\begin{eqnarray}}
\def\eea{\end{eqnarray}}

\begin{document}
\vspace*{4cm}
\title{NEW DIFFRACTION RESULTS FROM THE TEVATRON}

\author{ K. TERASHI\\
(for the CDF and D\O\hspace{1.4mm}Collaborations)}

\address{The Rockefeller University\\
1230 York Avenue, New York, NY 10021, USA}

\maketitle\abstracts{
We present new results from studies on diffractive dijet production and exclusive
production of dijet and diphoton obtained by the CDF Collaboration in proton-antiproton 
collisions at the Fermilab Tevatron.}

\section{Introduction}
Diffractive events in $\bar{p}p$ collisions are characterized by the presence 
of a leading proton or antiproton which remains intact, and/or a rapidity gap, 
defined as a pseudorapidity\footnote{The pseudorapidity $\eta$ of a particle 
is defined as $\eta \equiv -\ln (\tan \theta/2)$, where $\theta$ is the polar 
angle of the particle with respect to the proton beam direction.} region devoid 
of particles. Diffractive events involving hard processes (``hard diffraction''), 
such as production of high $E_T$ jets, 
have been studied to understand the QCD aspects of the exchanged 
object, the Pomeron (a color singlet entity with vacuum quantum numbers). 
One of the most important questions in hard diffractive 
processes is whether or not they obey QCD factorization, in other words, 
whether the Pomeron has a universal, process independent, parton 
distribution function (PDF). Results on diffractive deep inelastic scattering 
(DDIS) from the $ep$ collider HERA show that QCD factorization holds in DDIS. 
However, single diffractive (SD) rates of $W$-boson, dijet, $b$-quark and 
$J/\psi$ productions relative to non-diffractive (ND) ones measured in Run I 
at CDF \cite{CDF_Wjjb} are about an order of magnitude lower than expectations 
from PDFs determined at HERA, indicating a severe breakdown of QCD factorization 
in hard diffraction between Tevatron and HERA. 
The suppression relative to predictions based on DDIS 
PDFs has been further studied by measuring ``diffractive structure function'' 
$F_{jj}^D$ using diffractive dijet data in CDF \cite{CDF_SDjj_1800}.
The $F_{jj}^D$ was measured by looking at the ratio $R^{SD}_{ND}$
of SD to ND dijet event rates as a function of Bjorken-$x$ variable ($x_{Bj}$) 
and multiplying the ratio by the known ND proton structure function. 
Measuring the ratio $R^{SD}_{ND}$ thus provides information regarding the behavior of 
$F_{jj}^D$ relative to the ND proton PDF.

Another area of great interest is exclusive production of dijets, diphoton and
$\chi_{c(b)}$ meson at the Tevatron. In leading order (LO) QCD, such exclusive processes
can occur through exchange of a color-singlet two gluon system between nucleons, 
leaving large rapidity gaps in the forward regions. One of the two gluons takes 
part in a hard interaction while the other serves to neutralize the color. 
This type of production is generally suppressed by the Sudakov form factor. 
However, it's potentially a useful channel to search for the light Standard Model 
Higgs boson (predominantly decaying to $b\bar{b}$) at 
the LHC, since exclusive $b\bar{b}$ production is expected to be significantly 
suppressed by a helicity selection ($J_Z=0$) rule. Our goal of studies is 
to establish exclusive production experimentally and measure the 
cross sections of the exclusive processes to calibrate theoretical calculations
of exclusive Higgs production at the LHC.

\begin{figure}
\begin{center}
\psfig{figure=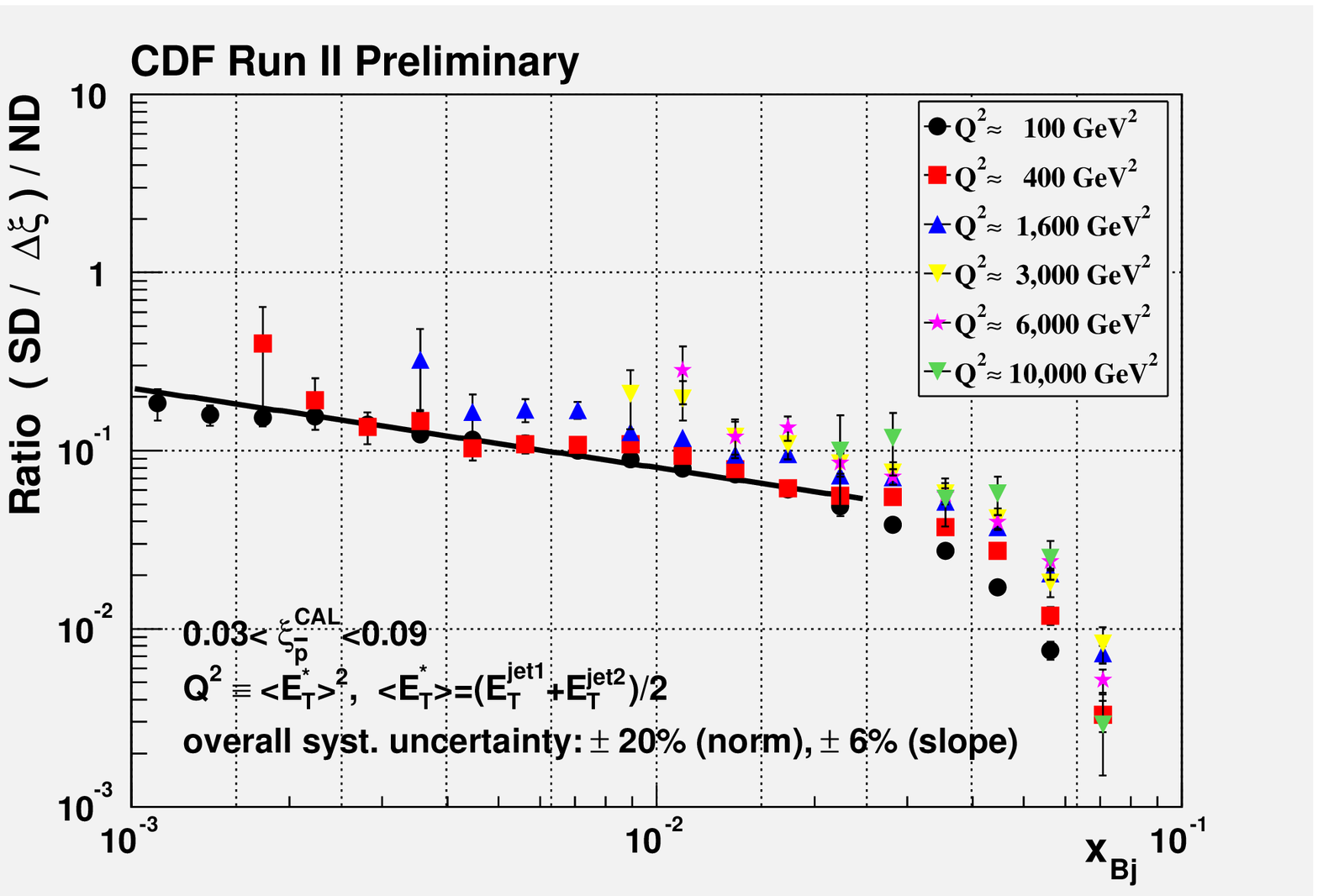,width=6cm}
\psfig{figure=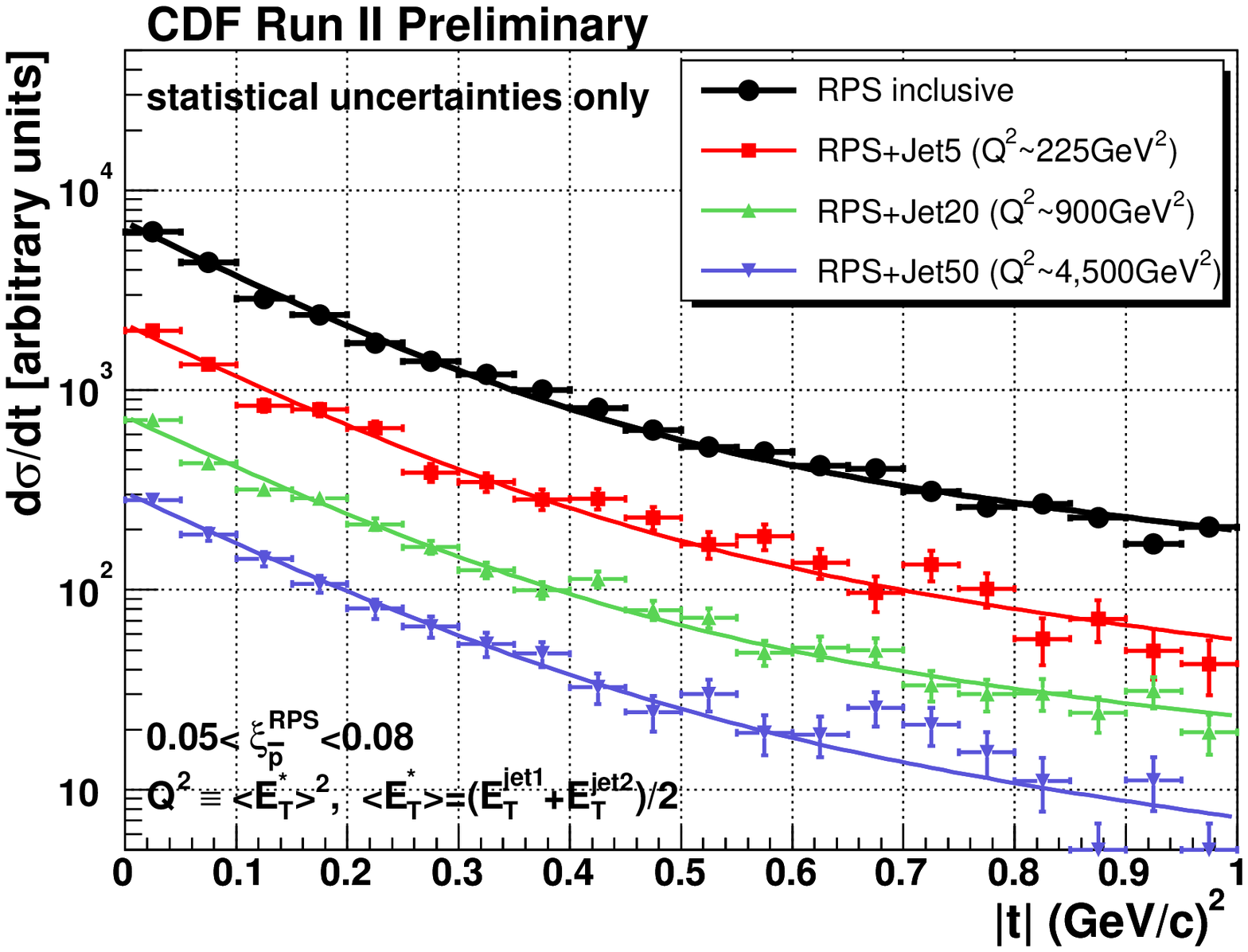,width=6cm}
\caption{{\it Left}: Ratio of SD to ND dijet rates as function of $x_{Bj}$ of the parton
in the antiproton for different $Q^2$ ranges. {\it Right}: $t$ distributions (with arbitrary
normalization) in diffractive dijet events with different $Q^2$ values.
\label{fig:jj}}
\end{center}
\end{figure}

\section{Run II Diffraction Measurements} 
In Run II, CDF has studied various topics on diffraction, 
including $Q^2$ dependence of $F_{jj}^D$ in SD and productions of exclusive dijet 
and exclusive diphoton, for which the results will be discussed below.
Two ``Miniplug'' (MP) calorimeters cover the forward pseudorapidity region 
$3.6<|\eta|<5.2$, and 7 stations of scintillation counters, called Beam Shower 
Counters (BSC), mounted around the beam pipe, extend the coverage to the 
very forward region of $5.4<|\eta|<7.4$. The Roman Pots (RP) used in Run I were 
re-installed and are being operated to trigger on leading antiprotons in the 
kinematic range $0.03<\xi<0.1$ and $0<|t|<3$ GeV$^2$, where $\xi$ is the fractional
momentum loss of the antiproton and $t$ is the four momentum transfer squared.

\section{Diffractive Dijet Production} 
Triggering on a leading antiproton in the RP in conjunction with at least one
jets in calorimeters, diffractive dijet events have been studied. 
Using a ND dijet sample triggered only on the jet requirement, the ratio $R^{SD}_{ND}$ 
is measured as a function of $x_{Bj}$, as shown in Fig.~\ref{fig:jj} (left).
This figure shows the ratios for different $Q^2$ values obtained from different
jet $E_T$ triggers. Here $Q^2$ is defined as the square of average value of the mean 
dijet $E_T$. In the range $100<Q^2<10000$ GeV$^2$ no significant $Q^2$ dependence is
observed, which indicates QCD evolution of the Pomeron could be similar to 
that of the proton. 

A $Q^2$ dependence of the $t$ in diffractive dijet 
events is also examined. Fig.~\ref{fig:jj} (right) shows the $t$ distributions 
(with arbitrary normalization) for different $Q^2$ values spanned over the wide
range. The slope at $t=0$ GeV/c$^{2}$ appears to be quite 
independent of $Q^2$ and is close to the one in standard diffractive $t$
distribution. Measurement of the slope values is currently under way.

\section{Search for Exclusive Dijet Production} 
We have implemented in Run II a dedicated trigger that requires a BSC gap on the
proton-side in addition to the requirements for the leading antiproton in the RP and
at least one calorimeter tower with $E_T>5$ GeV (RP+GAP+ST5). 
Requiring offline an additional gap in the MP on the $p$-side, 
we have obtained a large amount of double
Pomeron exchange (DPE) dijet events. For those events we examine ``dijet mass fraction'',
$R_{jj}$, defined as the invariant mass $M_{jj}$ of the two highest $E_T$ jets 
divided by the mass $M_X$ of the whole system (except leading nucleons);
$R_{jj}=M_{jj}/M_X$ \footnote{The system mass $M_X$ is obtained from 4-momenta of all 
calorimeter (massless) towers and the dijet mass $M_{jj}$ is calculated from the
calorimeter tower energies inside the R=0.7 cones of jets.}.
This observable should be sensitive to how much the event 
energy is concentrated in the dijet. The $R_{jj}$ of exclusive
dijet events is expected to be peaked around $R_{jj}\sim0.8$ and have a long
tail towards low $R_{jj}$ due to hadronization of partons causing energy leak from 
jet cones and also the presence of gluon radiations in the initial and final states.

\begin{figure}
\begin{center}
\psfig{figure=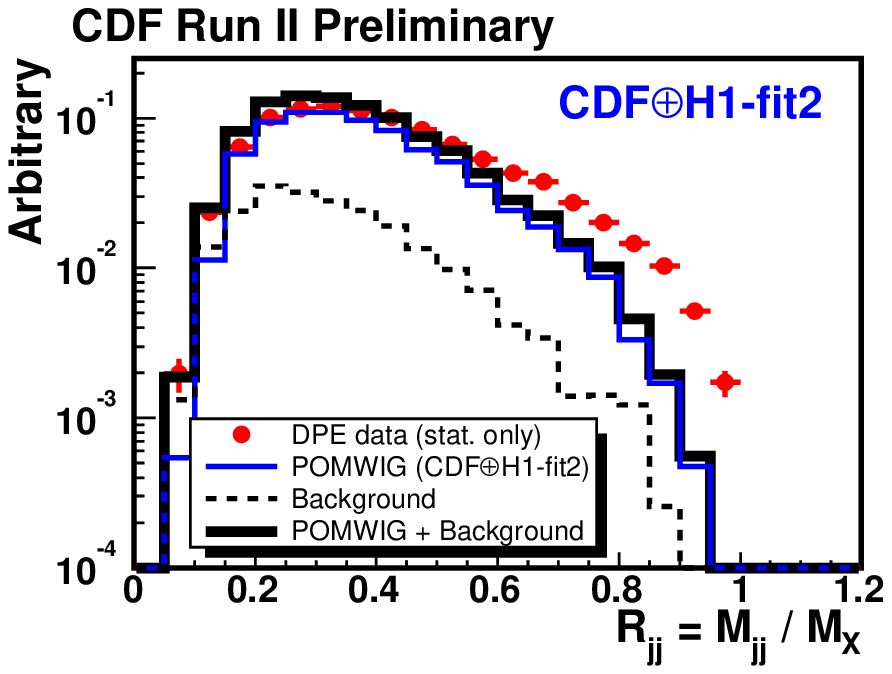,width=5.25cm}
\psfig{figure=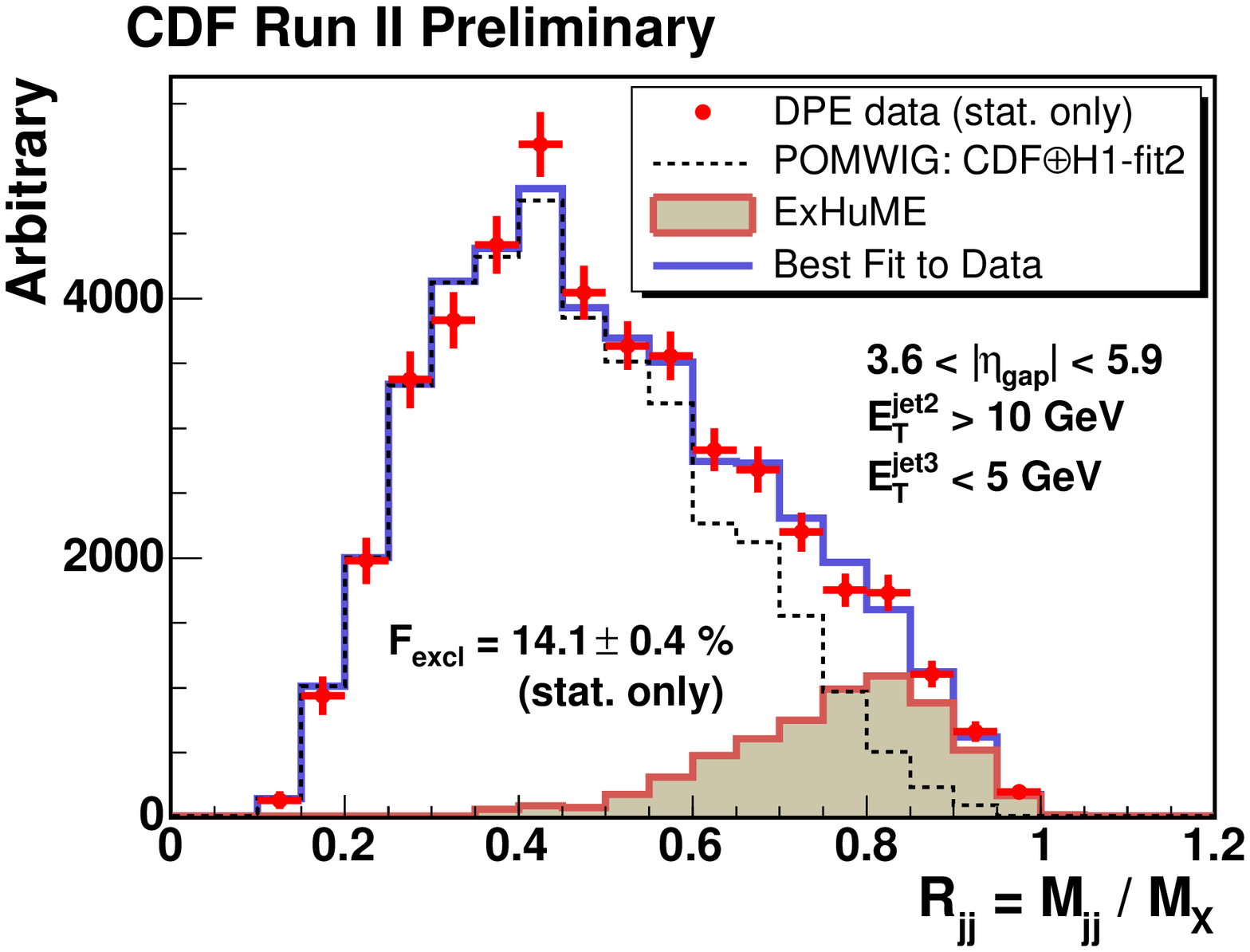,width=5.25cm}
\psfig{figure=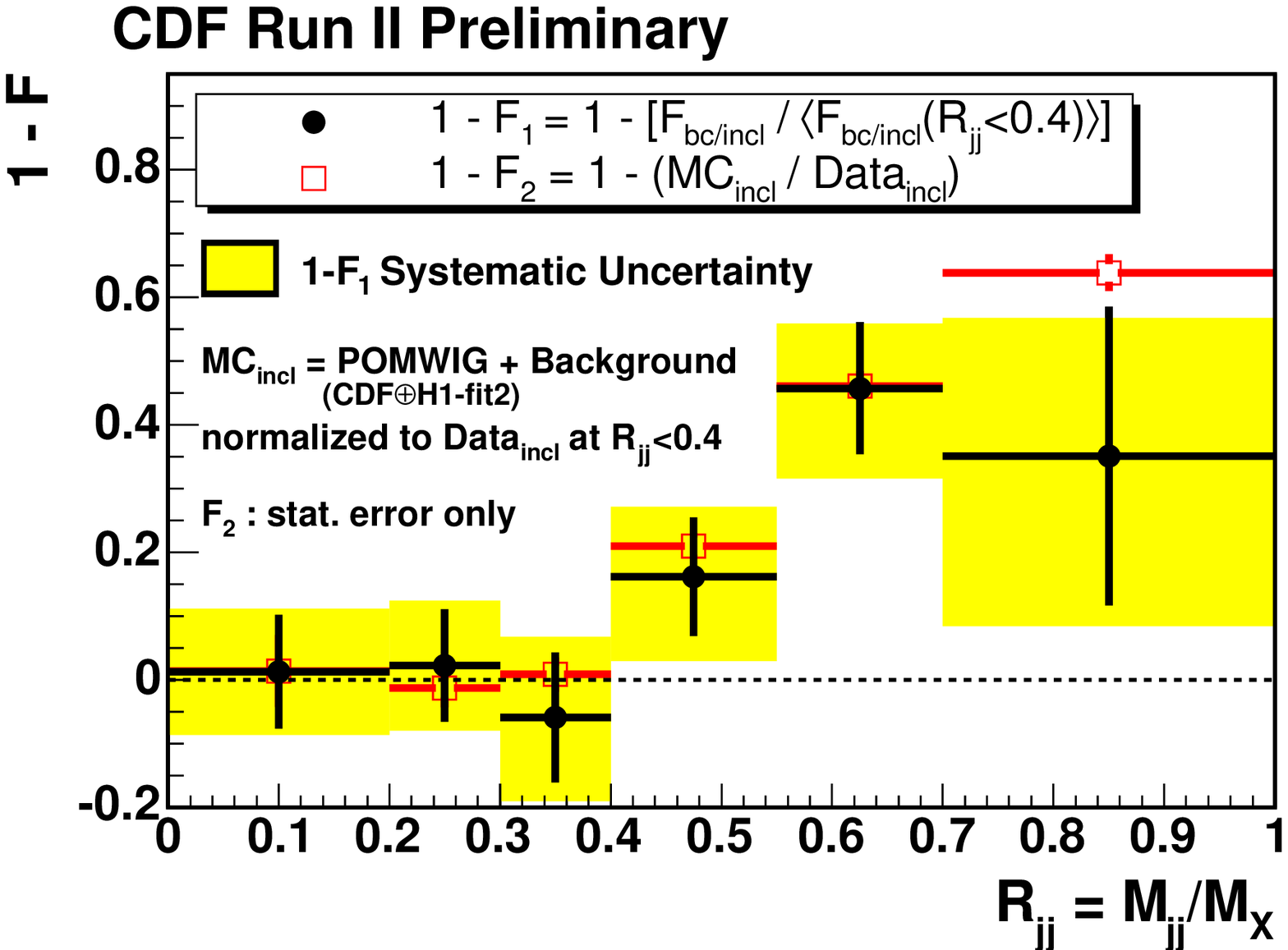,width=5.25cm}
\caption{{\it Left}: $R_{jj}$ of the data (points) and POMWIG MC prediction
(thick solid histogram) composed of DPE dijet events (thin solid histogram) and non-DPE
background events (dashed histogram). The data and the MC prediction are normalized to a same
area. {\it Center}: best fit (solid histogram) to the data (points) obtained 
using inclusive (dashed line) and exclusive ExHuME (shaded histogram) MC predictions for
the $R_{jj}$ shape. {\it Right}: One minus the ratio $F$, obtained from heavy flavor
jet ratio ($F_1$) and inclusive data-POMWIG comparison study ($F_2$), as a function of
$R_{jj}$. Details of the plot are described in the text.  
\label{fig:exc_jj}}
\end{center}
\end{figure}

The signal search is performed by comparing the data with predictions for inclusive
DPE dijet events in the $R_{jj}$ distribution shape and looking for an excess of
events in high $R_{jj}$ region. We use POMWIG Monte Carlo event generator \cite{POMWIG}
(with detector simulation) to simulate the DPE dijets. 
Fig.~\ref{fig:exc_jj} (left) shows a comparison of the $R_{jj}$ shapes between the data and 
the MC prediction composed of DPE dijet events and non-DPE background events. 
The data and the MC prediction are normalized to same area. We have examined various 
Pomeron PDFs and underlying events (Pomeron remnants) in the MC, and find that 
it is hard to reproduce the data excess at high $R_{jj}$ by any of those changes. 
Two exclusive dijet production models implemented in ExHuME \cite{ExHuME} and 
DPEMC \cite{DPEMC} (Exclusive DPE mode) have also been studied. Fig.~\ref{fig:exc_jj} 
(center) shows the best fit
to the data $R_{jj}$ shape obtained from the inclusive POMWIG and the exclusive ExHuME 
predictions for the $R_{jj}$ in events with dijets of $E_T>10$ GeV and a third jet
veto of $E_T^{jet3}<5$ GeV. The third jet veto is introduced because the exclusive MC
generates only a LO $gg \rightarrow gg$ process. The fit shows the data excess 
can be well described by the presence of exclusive dijets. The exclusive DPE
model in DPEMC also provides a good agreement with the data in the excess shape. 

According to Ref.~\cite{KMR_jj}, LO process of exclusive dijet is dominated by $gg \rightarrow gg$ 
and contributions from $gg \rightarrow q\bar{q}$ are strongly suppressed by the helicity 
selection rule. To check this prediction, we search for the suppression 
of heavy flavor ($b$ and $c$) jets in exclusive-dominant (high $R_{jj}$) region.
We use a 200 pb$^{-1}$ data sample triggered on the requirements of the RP+GAP+ST5 and
at least one displaced vertex track with $p_T>2$ GeV/c. The displaced track requirement effectively 
enhance heavy flavor contents in the sample. We measure the ratio $F$ of heavy flavor jets to all 
jets as a function of $R_{jj}$ of the events. The result is presented in Fig.~\ref{fig:exc_jj} 
(right) as ``$1-F$'' versus $R_{jj}$. 
The heavy flavor to all jet ratio ($F_1$ in the figure) is normalized by the weighted average of
the $F_1$ in the range $R_{jj}<0.4$ so that correlated systematic uncertainties are canceled out. 
We observe the increasing trend in $1-F_1$ with increasing $R_{jj}$, which could be a 
manifestation of the $J_Z=0$ selection rule.
The result is compared with the inclusive dijet result
by showing the ``$1-F_2$'', where $F_2$ is the ratio of the inclusive MC predicted events 
(normalized to the data at $R_{jj}<0.4$) to the data. The $1-F_2$ is thus 
equivalent to the fraction of the observed excess in the data.
The $1-F$ ratios are consistent with each other in both magnitude and $R_{jj}$ dependence. 

\section{Search for Exclusive Diphoton Production} 
\begin{figure}
\begin{center}
\psfig{figure=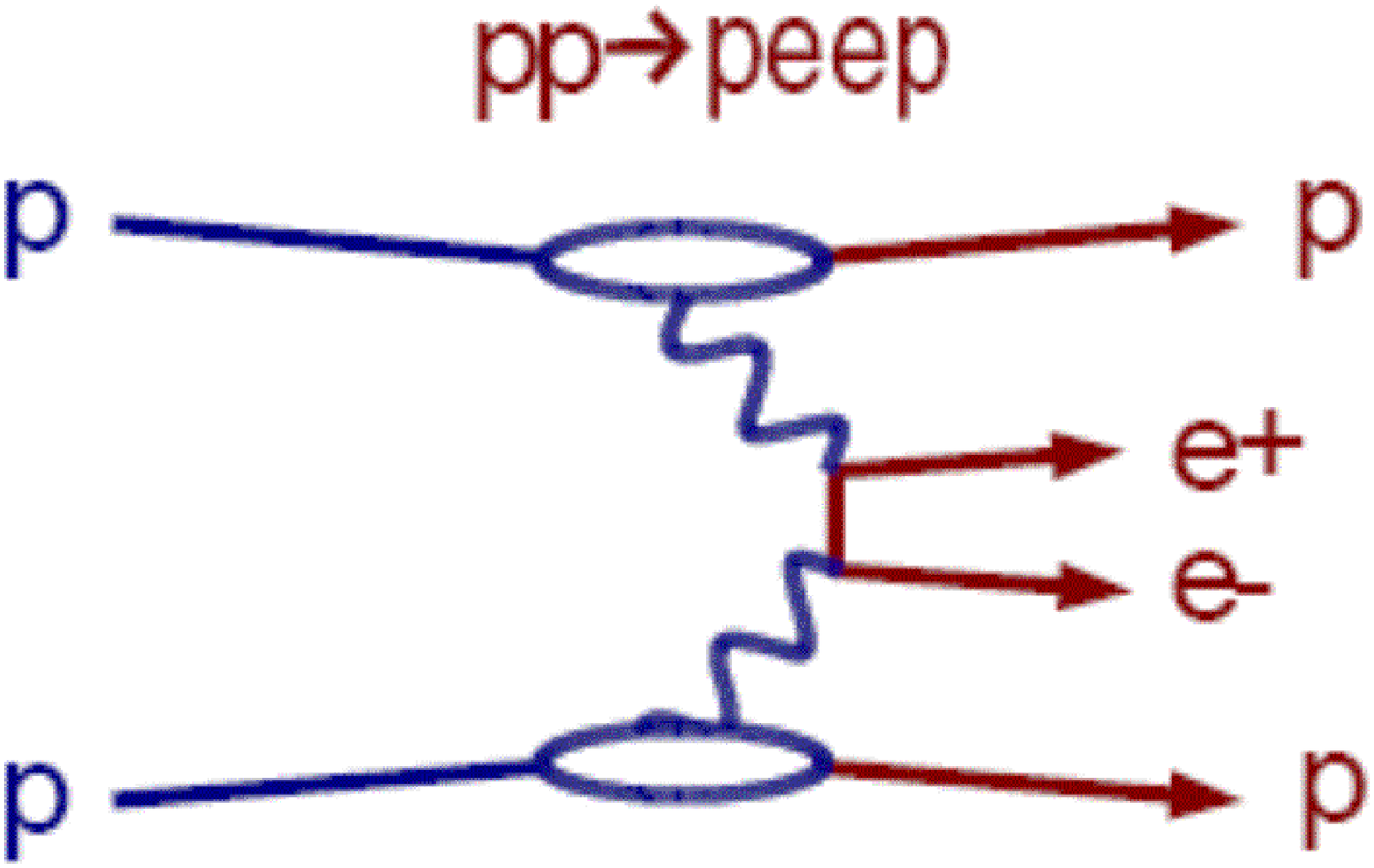,width=4.5cm}\hspace{1cm}%
\psfig{figure=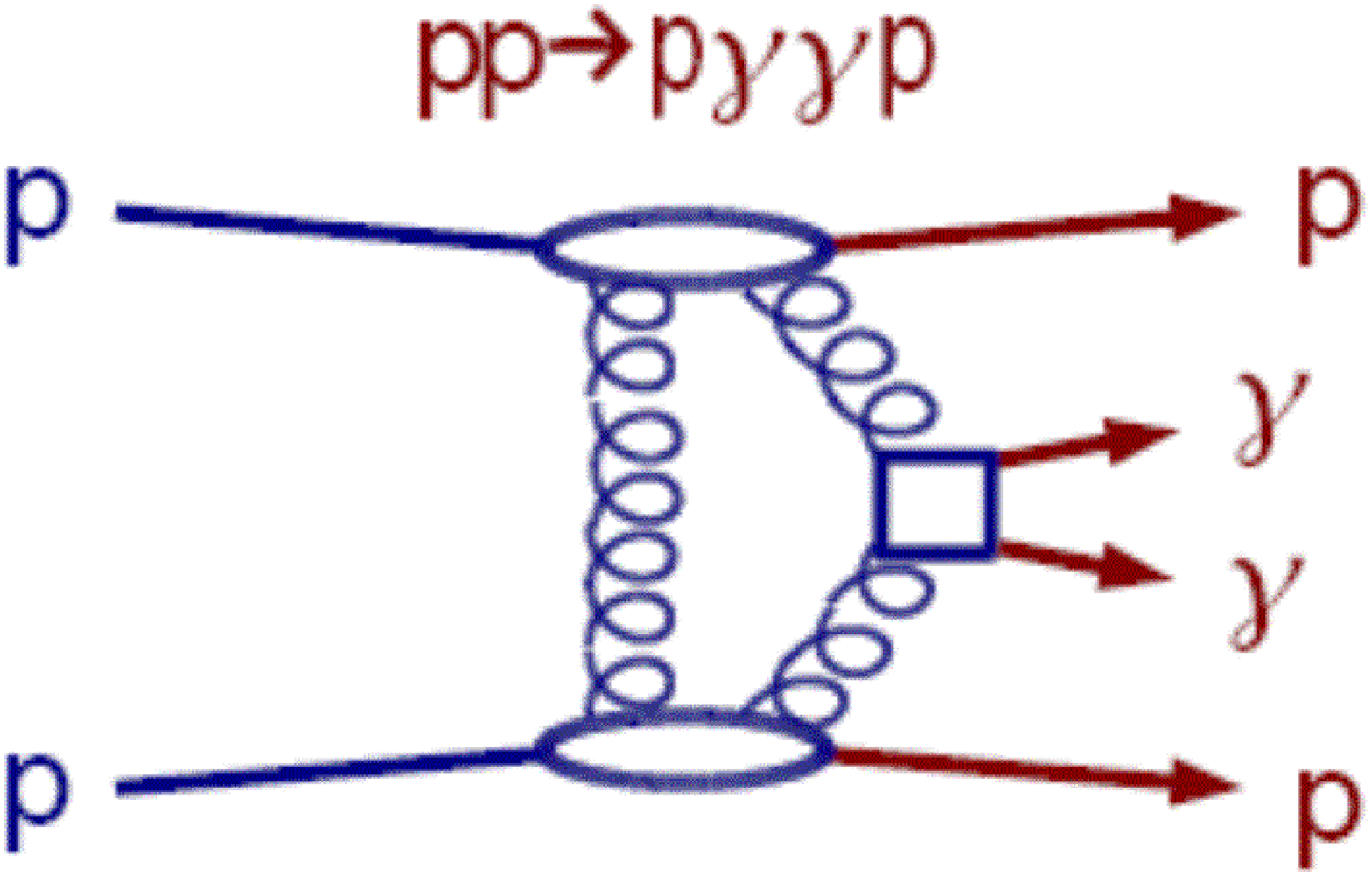,width=4.5cm}
\end{center}
\caption{{\it Left}: Exclusive production of $e^+e^-$ pair through two photon exchange 
$\gamma\gamma \rightarrow e^+e^-$ in $\bar{p}p$ collision. {\it Right}: Exclusive 
production of $\gamma\gamma$ pair via gluon-gluon fusion $gg \rightarrow \gamma\gamma$ 
with a quark loop in $\bar{p}p$ collision
\label{fig:exc_gg}}
\end{figure}

CDF has also performed search for exclusive diphoton production \cite{KMR_gg} 
(Fig.~\ref{fig:exc_gg}) using Run II data. The data used in the search is obtained by a 
trigger which requires two electromagnetic (EM) towers and BSC gaps in both forward directions. 
Requiring all the calorimeters to be empty above noise (except for the triggered two EM towers), 
we have observed 10 events containing two electron candidates with $E_T>5$ GeV 
(the two EM towers containing a single track with $p_T>1$ GeV/c each) and nothing else 
in the CDF detectors. The observed 
events appear to be consistent with QED-mediated dielectron production
$\bar{p}+p \rightarrow \bar{p}+e^+e^-+p$ through two photon exchange $\gamma\gamma \rightarrow e^+e^-$. 
The LPAIR Monte Carlo generator \cite{LPAIR} predicts $9\pm3$ events which are consistent with the 
observed events, though backgrounds in the data are not estimated yet.

In the same dataset the search finds 3 events with two $E_T>5$ GeV photon candidates
(the triggered EM towers associated with no tracks) and nothing else in the detectors. 
The ExHuME Monte Carlo generator for exclusive diphoton 
$\bar{p}+p \rightarrow \bar{p}+\gamma\gamma+p$ via $gg \rightarrow \gamma\gamma$ predicts 
$1^{+3}_{-1}$ events. Background estimation is currently under way.

\end{document}